\documentclass[osajnl,twocolumn,showpacs,superscriptaddress,10pt]{revtex4-1} 
\usepackage{amsmath,amssymb,graphicx}
\begin{document}

\title{Thermal effects on bipartite and multipartite correlations in fiber coupled cavity arrays}

\author{Jian-Song Zhang} \email{Corresponding author: jszhang1981@zju.edu.cn}
\affiliation{Department of Applied Physics, East China
Jiaotong University,
Nanchang 330013, People's Republic of China}

\author{Ai-Xi Chen}\email{Corresponding author: aixichen@ecjtu.jx.cn}
\affiliation{Department of Applied Physics, East China
Jiaotong University,
Nanchang 330013, People's Republic of China}

\begin{abstract}
We investigate the thermal influence of fibers on the dynamics of bipartite and multipartite correlations in
fiber coupled cavity arrays where each cavity is resonantly coupled to a two-level atom. The atom-cavity systems
connected by fibers can be considered as polaritonic qubits.
We first derive a master equation to describe the evolution of the atom-cavity systems.
The bipartite (multipartite) correlations is measured by concurrence and discord (spin squeezing).
Then, we solve the master equation numerically and study the thermal effects on the concurrence, discord, and spin squeezing of qubits.
On the one hand, at zero temperature, there are steady-state bipartite and multipartite correlations.
One the other hand, the thermal fluctuations of a fiber may blockade the generation of entanglement of two qubits connected directly by
the fiber while the discord can be generated and stored for a long time.
This thermal-induced blockade effects of bipartite correlations may be useful for quantum
information processing.
The bipartite correlations of a longer chain of qubits is more robust than a
shorter one in the presence of thermal fluctuations.
\end{abstract}


\maketitle 

\section{Introduction}

It is well-known that entanglement plays a fundamental role in quantum information
processing and quantum computation \cite{Nielsen2000}.
A quantum system is, in general, inevitably influenced by its surrounding environment \cite{Breuer2007}.
The entanglement dynamics of a quantum system
formed by several subsystems have been studied by many researchers \cite{Yu2004,Yu2009,Rau2008,Wu1996,Wu2001,Zhang2009,Zhang2010,Zhang2012}.
However, quantum entanglement is not the only kind of
quantum correlation useful for quantum information processing
\cite{Bennett1999,Horodecki2005,Niset2006,Modi2012,Ollivier2001,Henderson2001}. It was shown
both theoretically
\cite{Braunstein1999,Meyer2000,Datta2005,Datta2007,Datta2008,Dillenschneider2008,Sarandy2009,Cui2010}
and experimentally \cite{Lanyon2008} that some tasks can be sped up
over their classical counterparts using fully separable and highly
mixed states. These results show that separable states with
quantum discord may be useful for quantum information
processing. Quantum discord
\cite{Ollivier2001,Henderson2001} is another kind of quantum
correlation different from entanglement and has been investigated widely
\cite{Wang2010,Auyuanet2010,Sun2010,Werlang2009}.

In recent years, many efforts have been devoted to the study of coupled cavity array systems \cite{Serafini2006,Angelakis2007,Lijian2010,Umucalalar2012,Nissen2012,Chen2012,Jin2013,Kamide2013,Memarzadeh2011}.
In \cite{Serafini2006}, the authors proposed a scheme to implement effective quantum gates of two
two-level atoms within two distant cavities which are coupled by a fiber.
The photon-blockade-induced Mott transitions in coupled cavity arrays was studied \cite{Angelakis2007}.
It was shown that the four qubit W state and cluster state can be generated in a fiber coupled cavity array system \cite{Lijian2010}.
The fractional quantum Hall states of photons \cite{Umucalalar2012}, nonequilibrium dynamics \cite{Nissen2012},
polariton soliton \cite{Chen2012}, photon solid phases \cite{Jin2013}, and the first-order phase transition \cite{Kamide2013}
of coupled cavity arrays were investigated.
The authors of \cite{Angelakis2009} have studied a quantum system formed by two separate cavities coupled by a fiber or additional cavity
and shown that classical driving of the intermediate fiber or cavity can be used to create the entanglement between the
two ends in the steady state under dissipation. This is useful for the production of entanglement under realistic laboratory conditions.
In \cite{Memarzadeh2011}, the authors have studied the entanglement dynamics of coupled cavity arrays and shown
that the steady state entanglement can be achieved in the absence of thermal fluctuations.

In the present paper, we investigate the effects of thermal fluctuations of fibers on
the dynamics of bipartite and multipartite correlations
in fiber coupled cavity array systems.
Each cavity contains a two-level atom.  The atom-cavity systems connected by fibers can be regarded as
polaritonic qubits due to the blockade effect \cite{Serafini2006,Angelakis2007,Lijian2010}.
The bipartite correlations is measured by concurrence \cite{Wootters1998}
and discord \cite{Ollivier2001,Henderson2001}.
The multipartite correlations of the system is calculated by employing the spin squeezing \cite{Kitagawa1993,Ma2011}.
In order to describe thermal effects of the fibers on
the evolution of the atom-cavity systems (polaritonic qubits), we derive a master equation under the Markovian approximation.
Then, we solve the master equation numerically and study the thermal effects on the concurrence, discord, and spin squeezing of qubits.
Our results show that both the bipartite and multipartite correlations can be generated and stored for a long time at zero temperature.
One the other hand, the thermal fluctuations of a fiber could blockade the generation of entanglement
of two qubits which are connected directly by
the fiber. In contrast, the discord can be larger than zero. There is entanglement sudden death (ESD) phenomenon while
the discord vanishes only in the asymptotic limit, i.e., there is no sudden death of discord.
We also find that the bipartite correlations of a longer chain of qubits is more robust than a
shorter one if the thermal fluctuations of fibers is taken into considered.

The organization of this paper is as follows.
In section 2, we introduce the model and derive a master equation to describe
the fiber coupled cavity array system.
In section 3, we briefly review several measures of bipartite and multipartite correlations including
concurrence, discord and spin squeezing.
In section 4, we solve the master equation numerically and discuss the thermal effects of fibers on the
dynamics of bipartite and multipartite correlations.
In section 5, some conclusive remarks are given.

\section{The model and master equation}

\begin{figure}[tbp]
\centering {\scalebox{0.6}[0.7]{\includegraphics{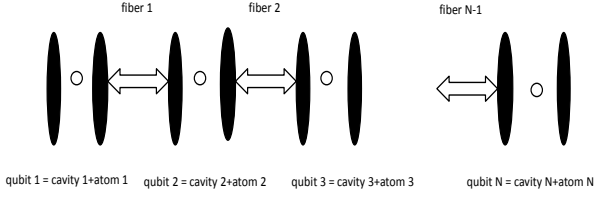}}}
\caption{The schematic picture of fiber coupled cavity arrays. This model consists of N coupled atom-cavity
system which are connected by N-1 fibers.
Each atom-cavity system can be treated as a
polaritonic qubit due to the blockade effect.
} \label{fig1}
\end{figure}

We consider a system formed by $N$ cavities connected by $N - 1$ fibers. Each cavity is resonantly coupled to a
two-level atom with frequency $\omega_0$. The schematic picture is shown in Fig.\ref{fig1}.
The Hamiltonian for the present system is \cite{Lijian2010,Angelakis2007}
\begin{eqnarray}
H &=& H_0 + H_{at-cav} + H_{cav-fib},\\
H_0 &=& \sum_{j} [\frac{\omega'_j}{2} (|e\rangle_j\langle e| - |g\rangle_j\langle g|) \\ \nonumber
&&+ \omega''_j  a^{\dagger}_j a_j]
+ \sum_{j,\alpha}\nu_{j\alpha}b_{j\alpha}^{\dagger}b_{j\alpha},  \\
H_{at-cav} &=& \sum_{j}^N g(a_j^{\dagger}|g\rangle_j\langle e| + h.c), \\
H_{cav-fib} &=& \sum_{j,\alpha}\xi_{j\alpha}(a_j^{\dagger}b_{j\alpha} + a_{j+1}^{\dagger}b_{j\alpha} + h.c),
\end{eqnarray}
with $h.c$ denoting the Hermitian conjugate. We assume the cavity is a quantized single-mode field;
$a^{\dagger}$ and $a$ are the creation and annihilation operators of the cavity field.
The subscripts $j$ and $\alpha$ stand for site $j$ and mode $\alpha$
of fiber $j$; $b_{j\alpha}^{\dagger}$ and $\nu_{j\alpha}$ are the creation operator and frequency of mode $\alpha$ in fiber $j$;
$g$ is the coupling strength between a cavity and the two-level atom within it;
$\xi_{j\alpha}$ is the coupling constant of cavity $j$ and mode $\alpha$ of fiber $j$;
 $|e\rangle_j$ and $|g\rangle_j$ are the excited and
ground states of two-level atoms. In the following, we assume the cavity and
two-level atom is coupled resonantly, and the frequencies of all cavities as well as atoms of different
sites are equal, i.e., $\omega_j' = \omega_j'' = \omega_0$.

The polaritonic states $|n, \pm\rangle_j$ are defined as
\begin{eqnarray}
|n, \pm\rangle_j  = \frac{1}{\sqrt{2}} (|n, g\rangle_j \pm |n-1, e\rangle_j),
\end{eqnarray}
with $|n\rangle_j$ denoting the Fock state of the j-th cavity.
It is worth noting that once a site of polariton is excited to $|1,-\rangle$
it is impossible to excite further due to the blockade effect \cite{Memarzadeh2011,Lijian2010,Angelakis2007}.
Consequently, we can treat each polariton as an effective two-level atom with
excited state $|E\rangle_j = \frac{1}{\sqrt{2}} (|1, g\rangle_j - |0, e\rangle_j)$
and ground state $|G\rangle_j = |g, 0\rangle_j$.
Therefore, the effective Hamiltonian of the present system is \cite{Memarzadeh2011,Lijian2010,Angelakis2007}
\begin{eqnarray}
H^{eff} &=& H_0^{eff} + H_{I}^{eff}, \\
H_0^{eff} &=& \frac{\omega}{2}\sum_{j}\sigma_{j}^{z}+\sum_{j,\alpha}\nu_{j\alpha}b_{j\alpha}^{\dagger}b_{j\alpha}, \nonumber\\
H_I^{eff} &=& \sum_{j,\alpha}(g_{j-1\alpha}\sigma_{j}^{-}b_{j-1\alpha}^{\dagger}+g_{j\alpha}\sigma_{j}^{-}b_{j\alpha}^{\dagger}+h.c),
\end{eqnarray}
with $\sigma_j^{+}=|E\rangle_j\langle G|$, $\sigma_j^{-} = (\sigma_j^+)^{\dagger} = |G\rangle_j\langle E|$,
$\sigma_j^{z} = |E\rangle_j\langle E| - |G\rangle_j\langle G|$,
$g_{j\alpha} = \xi_{j\alpha}/\sqrt{2}$ and $\omega = \omega_0 - g$.

Tracing out the degrees of freedoms of fibers, we have the following master equation of qubits in the interaction picture
(for more details, see Appendix A)
\begin{widetext}
\begin{eqnarray}
\frac{d\rho_{s}}{dt}&=& \mathcal{L} [\rho_s] = \sum_{j}\gamma_{j}n_{j}[2(\sigma_{j}^{+}+\sigma_{j+1}^{+})\rho_{s}(\sigma_{j}^{-}+\sigma_{j+1}^{-})
-(\sigma_{j}^{-}+\sigma_{j+1}^{-})(\sigma_{j}^{+}+\sigma_{j+1}^{+})\rho_{s}-\rho_{s}(\sigma_{j}^{-}
+\sigma_{j+1}^{-})(\sigma_{j}^{+}+\sigma_{j+1}^{+})] \nonumber\\
&&+\sum_{j}\gamma_{j}(n_{j}+1)[2(\sigma_{j}^{-}+\sigma_{j+1}^{-})\rho_{s}(\sigma_{j}^{+}
+\sigma_{j+1}^{+})-(\sigma_{j}^{+}+\sigma_{j+1}^{+})(\sigma_{j}^{-}+\sigma_{j+1}^{-})\rho_{s}-\rho_{s}(\sigma_{j}^{+}
+\sigma_{j+1}^{+})(\sigma_{j}^{-}+\sigma_{j+1}^{-})], \label{master_equation}
\end{eqnarray}
\end{widetext}
where $\mathcal{L}$ is a superoperator.
We note that in the case of $n_j = 0$ for all sites, the above master equation reduces to the master
equation of \cite{Memarzadeh2011}. The authors of \cite{Memarzadeh2011} have studied the steady state
entanglement in the absence of thermal fluctuations. In this work, we will investigated the thermal effects on
bipartite and multipartite correlations carefully by solving the above master equation Eq.(\ref{master_equation}) numerically.
For the sake of simplicity, we assume $\gamma_j = 1$ for all sites throughout this paper.

\section{Measures of bipartite and multipartite correlations}

Here, we briefly review some measures of bipartite and multipartite correlations
including concurrence, discord, and spin squeezing.

\subsection{Concurrence and discord}
The entanglement of a $2\times2$ quantum system which is described by density
matrix $\rho $ can be measured by concurrence which is defined as
\cite{Wootters1998}
\begin{equation}
C=\max {\{0,\lambda _{1}-\lambda _{2}-\lambda _{3}-\lambda _{4}\}},
\end{equation}%
where the $\lambda _{i}$ ($i=1,2,3,4$) are the square roots of the
eigenvalues in decreasing order of the magnitude of the \textquotedblleft
spin-flipped" density matrix operator $R=\rho (\sigma _{y}\otimes \sigma
_{y})\rho ^{\ast }(\sigma _{y}\otimes \sigma _{y})$ and $\sigma _{y}$ is the
Pauli Y matrix.

Quantum discord \cite{Ollivier2001,Henderson2001,Modi2012} is another kind of
quantum correlation which is different from entanglement.
For a bipartite system $\rho^{AB}$, the quantum mutual
information is defined as
\begin{eqnarray}
\mathcal{I}(\rho^{AB})&=&S(\rho^A)+S(\rho^B)-S(\rho^{AB}).\label{qmi1}
\end{eqnarray}
Here $S(\rho)=-Tr(\rho
\log_2{\rho})=-\sum_i(\lambda_i\log_2{\lambda_i})$ is the von
Neumann entropy of density matrix $\rho$ with $\lambda_i$ being the
eigenvalues of density matrix $\rho$; $\rho^A$($\rho^B$) is the reduced density matrix
of $\rho^{AB}$ by tracing out subsystem $B(A)$.
To quantify quantum discord, the authors of \cite{Ollivier2001}
proposed to employ the von Neumann type measurements consisting of
one-dimensional projector $\{\mathcal{B}_i\}$ which acts on system $B$
only with $\sum_i\mathcal{B}_i=1$. The conditional density
matrix of the total system
 after the von Neumann type measurements is \cite{Ollivier2001}
\begin{eqnarray}
\rho^{AB}_{\mathcal{B}_i}&=&\frac{1}{p_i}(I\otimes\mathcal{B}_i )\rho^{AB}(I\otimes\mathcal{B}_i ),\nonumber\\
p_i&=&Tr((I\otimes\mathcal{B}_i )\rho^{AB}(I\otimes\mathcal{B}_i )),
\end{eqnarray}
where $p_i$ is the probability of the corresponding measurement. The
quantum conditional entropy with respect to this kind of measurement
is defined as
\begin{eqnarray}
S(\rho^{AB}{|\{\mathcal{B}_i\}})=\sum_i p_i S(\rho^{AB}_{\mathcal{B}_i}). \label{qce1}
\end{eqnarray}
The corresponding quantum mutual information with respect to the
measurement is defined by
\begin{eqnarray}
\mathcal{I}(\rho^{AB}|\{\mathcal{B}_i\})=S(\rho^A)-S(\rho^{AB}|\{\mathcal{B}_i\}).
\end{eqnarray}
Here $\mathcal{I}(\rho^{AB}|\{\mathcal{B}_i\})$ is the
information obtained about system $A$ after one performs measurement
$\mathcal{B}_i$ on subsystem $B$. The classical correlation
 is defined as \cite{Ollivier2001,Henderson2001}
\begin{eqnarray}
\mathcal{J}(\rho^{AB})&=&\sup_{\{\mathcal{B}_i\}}\mathcal{I}(\rho^{AB}|\{\mathcal{B}_i\})\nonumber\\
&=&S(\rho^A)-\min_{\{\mathcal{B}_i\}} [S(\rho^{AB}|\{\mathcal{B}_i\})]. \label{cc1}
\end{eqnarray}
The quantum discord is obtained by subtracting $\mathcal{J}$ from
the quantum mutual information $\mathcal{I}$
\begin{eqnarray}
\mathcal{D}(\rho^{AB})=\mathcal{I}(\rho^{AB})-\mathcal{J}(\rho^{AB}). \label{qd1}
\end{eqnarray}

\subsection{Spin squeezing}
In order to study multipartite quantum correlations, we adopt the measure spin squeezing which was introduced by
Kitagawa and Ueda \cite{Kitagawa1993}. For a more recent review of spin squeezing, see Ref.\cite{Ma2011}.
The main advantage of spin squeezing as a measure of multipartite correlations is that
it is relatively easy to generate and measure spin squeezing experimentally since spin-squeezing parameters only
involve the first and second moments of the collective angular momentum operators.
The spin squeezing is defined by \cite{Kitagawa1993}
\begin{eqnarray}
\xi_s^2 = \frac{4(\Delta J_{\bot})^2_{min}}{N},
\end{eqnarray}
where $N$ is the number of particles and the minimization in the above equation is taken over all
directions denoted by $\bot$, which are perpendicular to the mean
spin direction $\langle \overrightarrow{J}\rangle /|\langle \overrightarrow{J}\rangle |$.
After some algebra, the spin squeezing can be derived as \cite{Ma2011}
\begin{eqnarray}
\xi_s^2 &=& \frac{2}{N}[ \langle(\overrightarrow{J}_{\overrightarrow{n_1}}^2 + \overrightarrow{J}_{\overrightarrow{n_1}}^2)\rangle\nonumber\\
&&- \sqrt{ \langle(\overrightarrow{J}_{\overrightarrow{n_1}}^2 - \overrightarrow{J}_{\overrightarrow{n_1}}^2)\rangle
+4 cov(\overrightarrow{J}_{\overrightarrow{n_1}}, \overrightarrow{J}_{\overrightarrow{n_2}})}], \nonumber\\
\overrightarrow{n_1} &=& (-\sin \phi, \cos \phi, 0), \nonumber\\
 \overrightarrow{n_2} &=& (\cos \theta \cos \phi, \cos \theta \sin \phi, -\sin \theta), \nonumber \\
\theta &=& \arccos \frac{\langle J_z \rangle }{|\overrightarrow{J}|}, \nonumber\\
cov(x, y) &=& \frac{1}{2}(\langle x y\rangle + \langle y x\rangle ) - \langle x\rangle \langle y \rangle,
\end{eqnarray}
with
\begin{eqnarray}
\phi = \left\{ \begin{array}{cc}
 \arccos\frac{\langle J_x \rangle }{|\overrightarrow{J}|\sin\theta},         & \langle J_y\rangle > 0, \\
2\pi -  \arccos\frac{\langle J_x \rangle }{|\overrightarrow{J}|\sin\theta},  & \langle J_y\rangle \leq 0.
\end{array} \right.
\end{eqnarray}
In the case of $\xi_s^2 < 1$, there is multipartite quantum correlations.
Spin squeezing is naturally connected with quantum correlations such as entanglement \cite{Ma2011}.
It has a very close relation with concurrence and is significant for entanglement detection.
But, spin squeezing and pairwise entanglement are different.
In fact, a spin squeezed state with $\xi_s^2 < 1$ is pairwise entangled,
while a pairwise entangled state may not be spin-squeezed according to the squeezing parameter
 (see page 116 of \cite{Ma2011}).

\section{Thermal effects on dynamics of correlations}

\begin{figure}[tbp]
\centering {\scalebox{0.38}[0.5]{\includegraphics{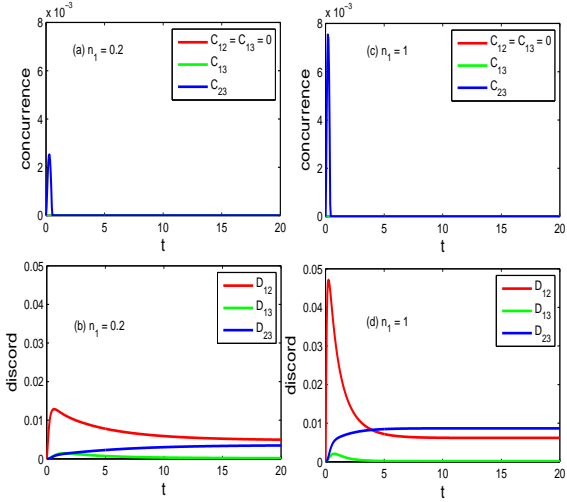}}}
\caption{The concurrence and discord of two qubits are plotted as functions of time t for $n_1 = 0.2$ and $n_1 = 1$ with
$n_2 = 0$, $N = 3$, and the initial state is $|\psi(0)\rangle  = |GGG\rangle$.
} \label{fig2}
\end{figure}

\begin{figure}[tbp]
\centering {\scalebox{0.38}[0.5]{\includegraphics{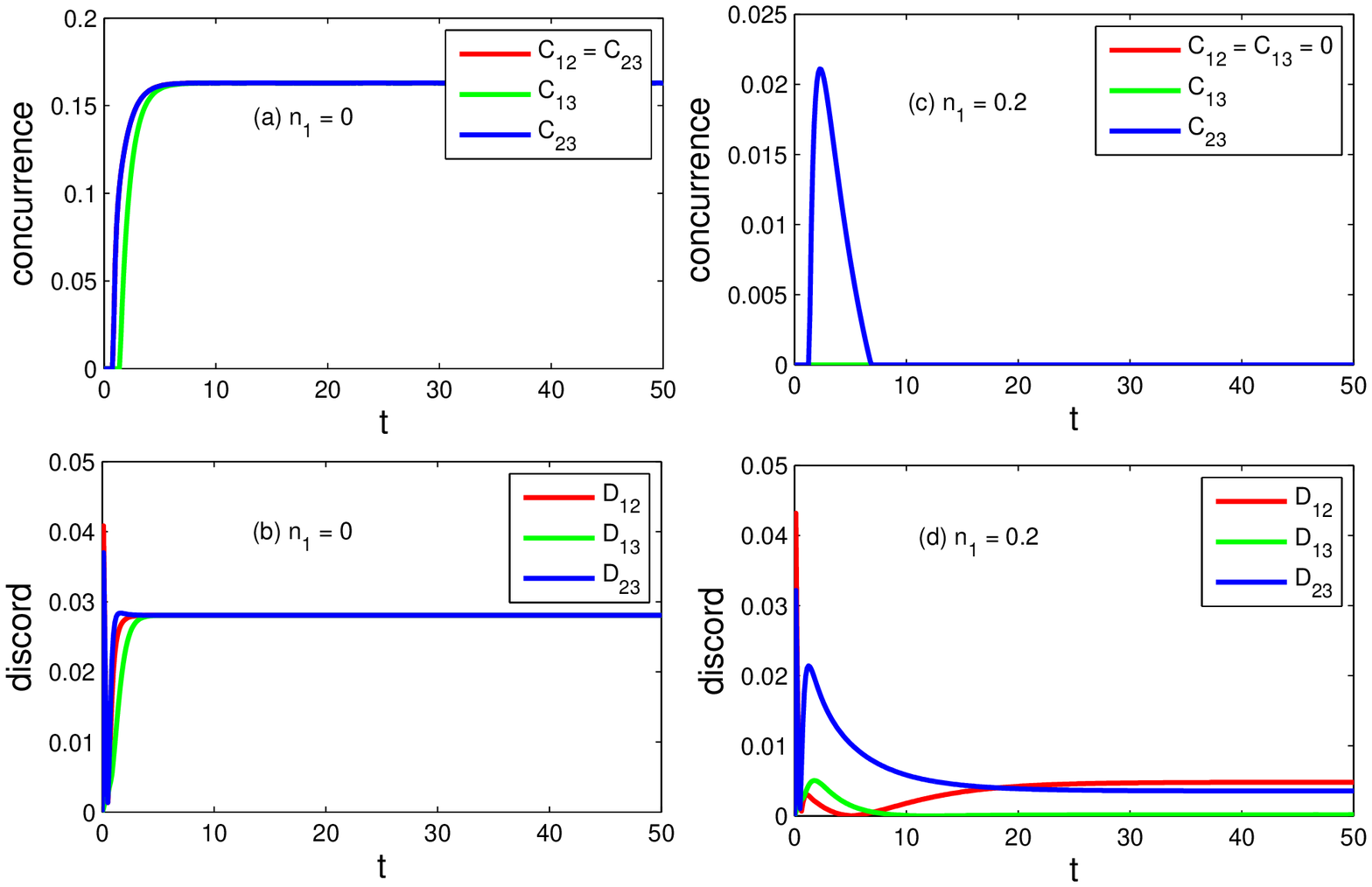}}}
\caption{The concurrence and discord of two qubits are plotted as functions of time t for $n_1 = 0$ and $n_1 = 0.2$ with
$n_2 = 0$, $N = 3$, and the initial state is $|\psi(0)\rangle  = |EEE\rangle$.
} \label{fig3}
\end{figure}

\begin{figure}[tbp]
\centering {\scalebox{0.42}[0.55]{\includegraphics{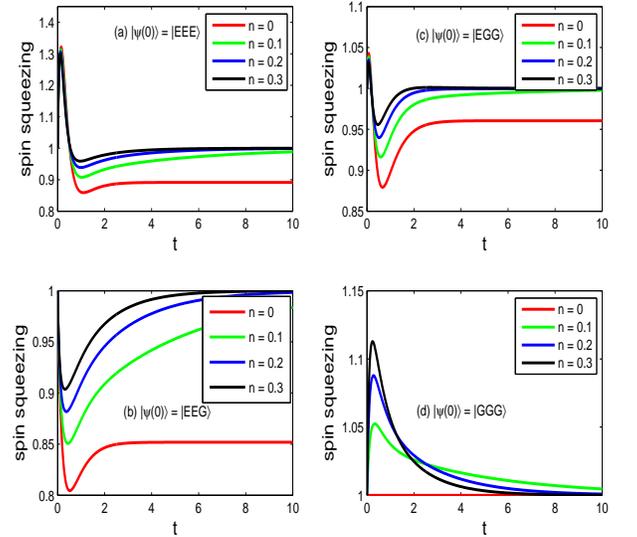}}}
\caption{The spin squeezing of the sytem are plotted as functions of time t for different
initial states with $N = 3$ and $n_1 = n_2 = n$.
} \label{fig4}
\end{figure}

In this section, we solve the master equation numerically and calculate bipartite and multipartite correlations
of the fiber coupled cavity arrays using concurrence, discord, and spin squeezing.
At zero temperature, if all the qubits are initially prepared in ground states
$|\Psi_G\rangle = |GG\cdots G\rangle = |G\rangle_1|G\rangle_2\cdots |G\rangle_N $,
then the system will not evolve due to the fact that
$\mathcal{L} [|\Psi_G\rangle\langle \Psi_G|]= 0$ (see Eq.(\ref{master_equation})).
Therefore, no bipartite and multipartite correlations will be generated if all the qubits are initially prepared in ground states
when the fibers are at zero temperature.

\subsection{Bipartite correlations}
\subsubsection{Ground states}
Now, we consider the influence of thermal fluctuations on bipartite correlations for different initial states.
If we increase the temperature of fibers, then there is bipartite correlations when
the qubits are initially prepared in ground states $|\Psi_G\rangle$ as one can see
from Fig.\ref{fig2}. We assume $N = 3$, $n_1 >0 $, and $n_2 = 0$ in Fig.\ref{fig2}.
The concurrence of qubits 2 and 3 denoted by $C_{23}(t)$ first increases with time and then reaches a maximal value.
Comparing Fig.\ref{fig2}(a) and Fig.\ref{fig2}(c), one can find the maximal value of $C_{23}$ will
be enhanced by increasing the temperature of fibers. 
As the system evolves, qubits 2 and 3 become separable eventually, i.e.,
there is ESD phenomenon \cite{Yu2004,Yu2009}.
This implies that the thermal fluctuations of
environments may play a constructive role in the present model. However, 
we should note that the entanglement of qubits will be destroyed when the temperature 
is high enough. In the case of $n_1 = n_2 = 0.2$ there is no entanglement of qubits.
However, there is steady sate discord of qubits as one can clearly see from Fig.\ref{fig2}(b) and Fig.\ref{fig2}(d).
It is worth noting that the concurrence of qubits 1 and 2 ($C_{12}$) as well as 1 and 3 ($C_{13}$)
are always zero. This indicates that the thermal fluctuations of fiber 1 which connects qubits 1 and 2 as well as qubits 1 and 3 will
prevent the generation of entanglement between these qubits.

\subsubsection{Excited states}
In Fig.\ref{fig3}, we plot concurrence and discord of two qubits for $n_1 = 0$ ((a) and (b)) and $n_1 = 0.2$ ((c) and (d))
with $N = 3$, $n_2 = 0$, and $|\psi(0)\rangle = |EEE\rangle$. From Fig.\ref{fig3}(a), we find
the entanglement of two qubits does not increase smoothly, i.e., it suddenly becomes
larger than zero as the system evolves. This phenomenon is called entanglement sudden birth (ESB) \cite{Ficek2008}.
Compared to qubits 1 and 2 (or qubits 2 and 3), it takes a longer time for qubits 1 and 3 to become entangled.
In the case of $n_1 = 0$, the steady state entanglement or discord of qubits are the same.
If we increase the temperature of fiber 1, there is ESD of qubits 2 and 3. Note that qubits 1 and 2 (or qubits 1 and 3)
are always separable, i.e., the thermal fluctuations of fiber 1
will blockade the generation of entanglement of qubits 1 and 2 as well as qubits 1 and 3
which are connected by fiber 1. In the case of $n_1 = 0.2$, there is steady state
discord. Different from the case of $n_1 = 0$, the values of steady state discord are different for different qubits.

\begin{figure}[tbp]
\centering {\scalebox{0.5}[0.6]{\includegraphics{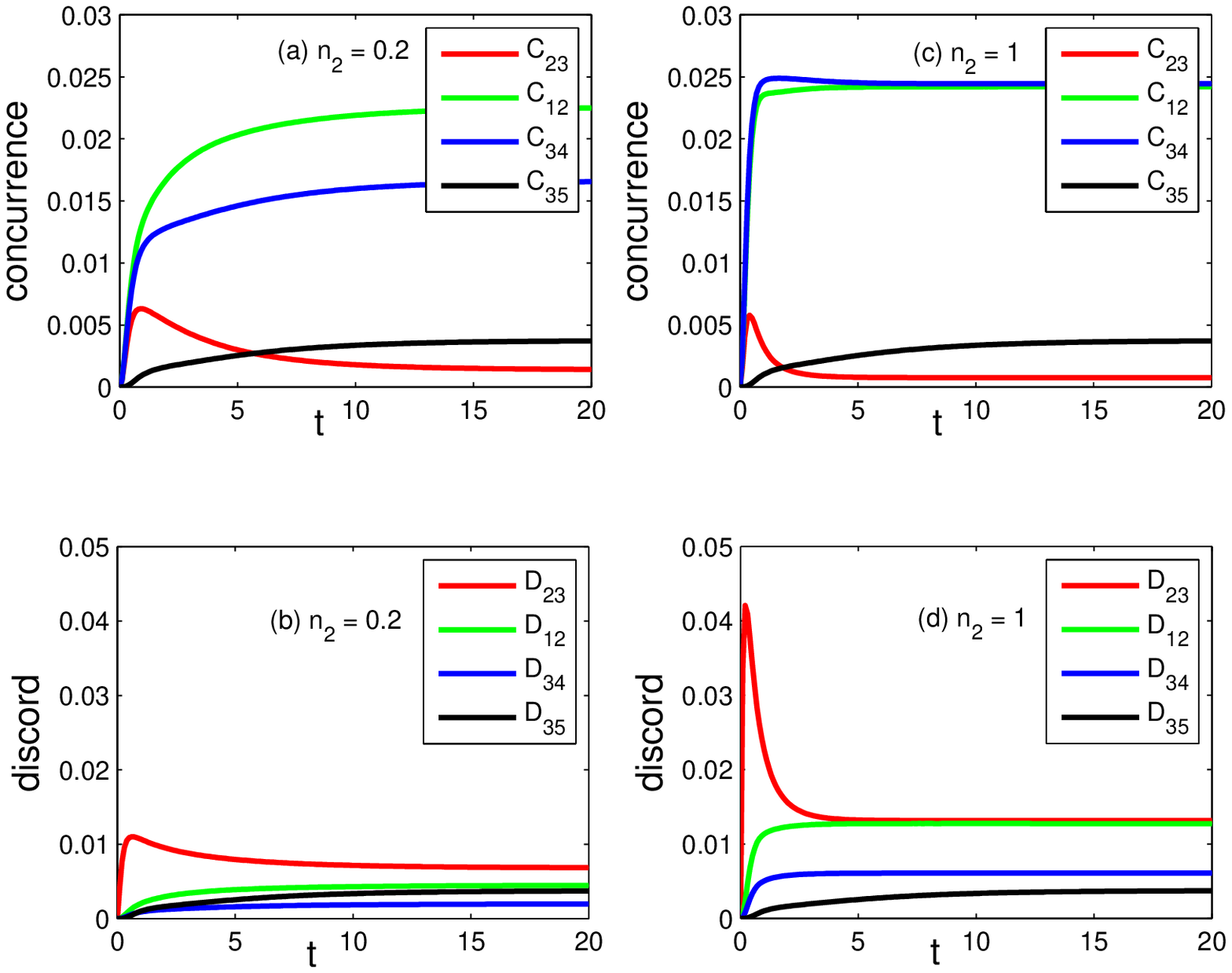}}}
\caption{The concurrence and discord of two qubits are plotted as functions of time t for $n_2 = 0.2$ and $n_2 = 1$ with
$n_1 = n_3 = n_4 = 0$, $N = 5$, and the initial state is $|\psi(0)\rangle  = |GGGGG\rangle$.
} \label{fig5}
\end{figure}

\begin{figure}[tbp]
\centering {\scalebox{0.5}[0.6]{\includegraphics{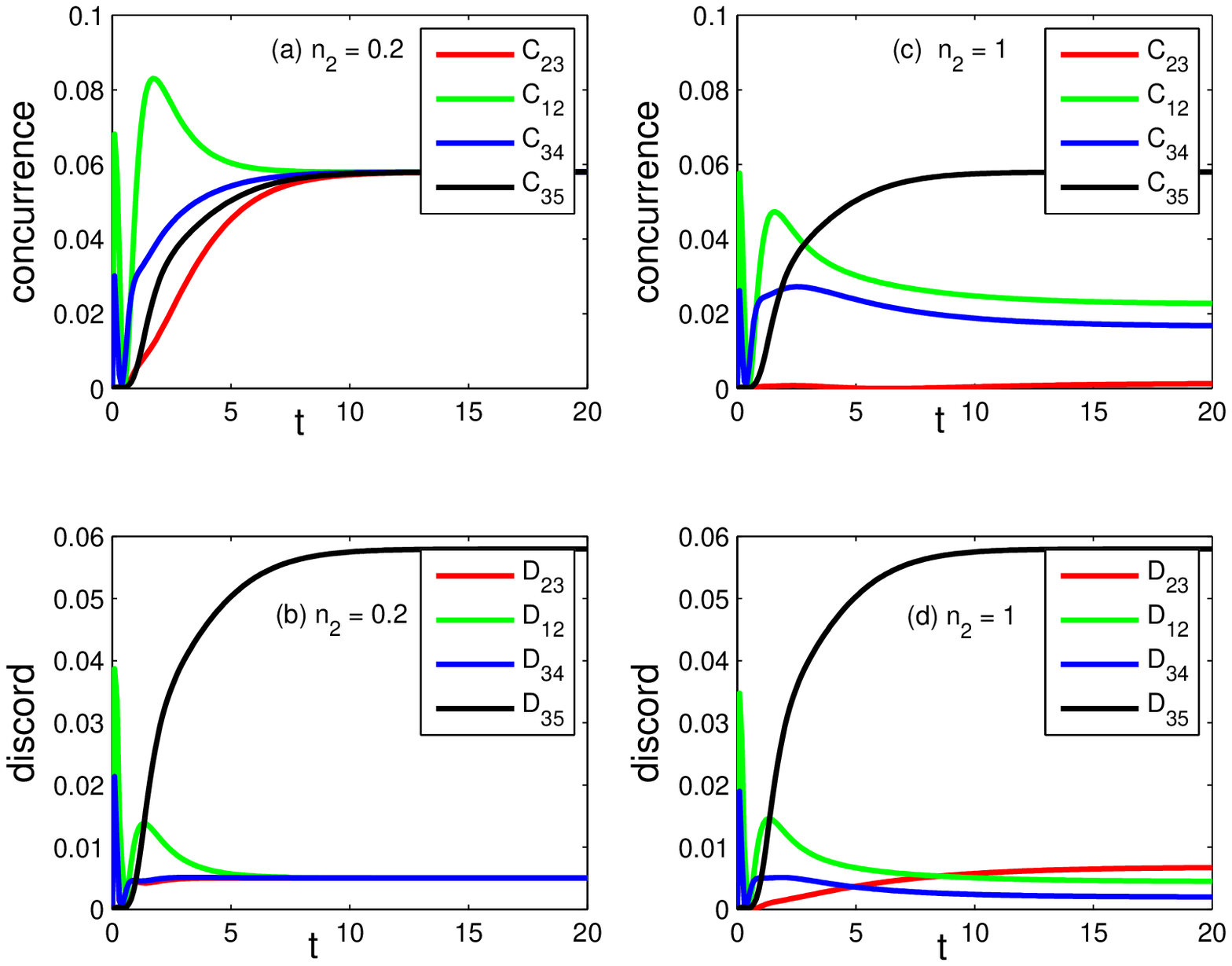}}}
\caption{The concurrence and discord of two qubits are plotted as functions of time t for $n_2 = 0.2$ and $n_2 = 1$ with
$n_1 = n_3 = n_4 = 0$, $N = 5$, and the initial state is $|\psi(0)\rangle  = |EEEEE\rangle$.
} \label{fig6}
\end{figure}

\subsection{Multipartite correlations}
The multipartite correlations is measured by spin squeezing $\xi_s^2$ introduced by
Kitagawa and Ueda \cite{Kitagawa1993}. In Fig.\ref{fig4}, we plot spin squeezing of
qubits as functions of time for different initial states and several average number of photons of fibers $n$ with $N = 3$.
If all the fibers are in zero temperature, then the spin squeezing $\xi_s^2$ can always be smaller than 1 which implies
there is multipartite correlations of qubits in the absence of thermal fluctuations.
When the initial state is $|GGG\rangle$, the spin squeezing of the system is larger or equal to 1.
The dynamics of spin squeezing for different initial states are different.
For instance, if the initial state is $|EEE\rangle$, the spin squeezing first increases with time (lager than 1)
and then reaches the maximal value. Finally, it will be a constant.
However, if the system is initially prepared in state $|EEG\rangle$, the spin squeezing
first decreases with time and then reaches the minimal value. Most of time, the spin
squeezing is smaller than 1 which indicates that there is multipartite correlations if the initial state is
$|EEG\rangle$ as one can see from Fig.\ref{fig5}(b).

\subsection{Influence of the number of qubits}
We now turn to discuss the influence of the number of qubits on the dynamics of quantum correlations.
As we have noted previously, for a system formed by 3 qubits, the entanglement of two qubits which are not connected
directly by fibers can only exist for a short time, i.e., the entanglement of two qubits
will be completely destroyed by the thermal fluctuations of fibers eventually. However,
there is steady state entanglement for a system of 5 qubits as one can see from Figs.\ref{fig5} and \ref{fig6}.
In Fig.\ref{fig5}, the initial state is $|GGGGG\rangle$ and $n_1 = n_3 = n_4 = 0$. The temperature of
fiber 2 is larger than zero. Comparing the red lines of Fig.\ref{fig5}, we see that the influence of thermal fluctuations
of fiber on the dynamics of concurrence and discord is very different. In Fig.\ref{fig5}(a) and Fig.\ref{fig5}(c),
the steady state concurrence $C_{23}$ is the minimal of all concurrence. However,
the steady state discord $D_{23}$ is the maximal one. Therefore, systems with more entanglement may
have less discord.

\section{Conclusions}
In the present paper, we have investigated the influence of thermal fluctuations of fibers on the
bipartite and multipartite correlations of fiber coupled cavity array systems.
Each cavity is resonantly coupled to a two-level atom.
The atom-cavity systems which are connected by fibers can be considered as polaritonic qubits due to the blockade effect.
First, we derived a master equation to describe the dynamics of the model by using the standard techniques of quantum
theory. In the absence of thermal fluctuations of fibers,
this master equation reduces to the previous results \cite{Memarzadeh2011}.
The master equation was solved numerically and thermal effects of fibers
on the concurrence, discord, and spin squeezing of qubits
was discussed.
At zero temperature, the entanglement, discord, and spin squeezing can be preserved for a long time.
The thermal fluctuations of fibers destroys the steady state correlations.
Eventually, two qubits becomes completely separable while the discord of them vanishes
only asymptotically. In particular, the thermal fluctuations of a fiber may blockade the generation of entanglement of
two qubits connected directly by the fiber. Evidently, the discord behaves very differently from the entanglement
since the discord can be generated and stored for a long time.
This thermal-induced blockade effects of bipartite correlations allows us to prevent the generation of one kind of
quantum correlations (entanglement) while permitting the generation of others such as discord.
We also discuss the influence of the number of sites on the dynamics of quantum correlations and found that
the bipartite correlations of a longer chain of qubits is more robust than a
shorter one in the presence of thermal fluctuations.
In addition, we found that systems with more entanglement may have less discord.

\section*{Acknowledgement}
This work is supported by
National Natural Science Foundation of China with Grant Nos. 11047115, 11365009 and 11065007,
the Scientific Research Foundation of Jiangxi
with Grant No. 20122BAB212008,
and the Scientific Research Foundation of Jiangxi
Provincial Department of Education with Grant No. GJJ12294.

\appendix

\section{Derivation of master equation}
Here, we derive the master equation of Eq.(\ref{master_equation}).
The reduced density matrix of
the system in the interaction picture obeys the following master equation(we have set $\hbar = 1$) \cite{Scully1997}
\begin{widetext}
\begin{eqnarray}
\frac{d\rho_{s}(t)}{dt} = -i Tr_{B}[V(t),\rho_{s}(0)\otimes\rho_{B}]
-Tr_{B}\int_{0}^{\infty}dt'[V(t),[V(t'),\rho_{s}(t')\otimes\rho_{B}]]. \label{reduced_density_matrix}
\end{eqnarray}

Note that $\rho_B$ is the density matrix of the bath. In the present model,
we treat the fibers as the thermal bath. The interaction Hamiltonian in the interaction picture is
\begin{eqnarray}
V(t)&=&e^{iH_{0}^{eff}t}H_{I}^{eff}e^{-iH_{0}^{eff}t}
= \sum_{j,\alpha}(e^{-i\Delta_{j-1\alpha}t}g_{j-1\alpha}\sigma_{j}^{-}b_{j-1\alpha}^{\dagger}
+e^{-i\Delta_{j\alpha}t}g_{j\alpha}\sigma_{j}^{-}b_{j\alpha}^{\dagger} + h.c),
\end{eqnarray}
with $\Delta_{j\alpha} = \omega - \nu_{j\alpha}$.
Using the relations \cite{Scully1997,Ficek2008}
\begin{eqnarray}
\langle b_{j\alpha}^{\dagger}b_{j'\alpha'}\rangle&=&n_{j}\delta_{jj'}\delta_{\alpha\alpha'}, \\
\langle b_{j\alpha}b_{j'\alpha'}^{\dagger}\rangle&=&(n_{j}+1)\delta_{jj'}\delta_{\alpha\alpha'}, \\
\sum_{\alpha}|g_{j\alpha}|^{2}e^{\pm i\Delta_{j\alpha}t}&=&\sum_{\alpha}|g_{j\alpha}|^{2}e^{\pm i(\omega-\nu_{j\alpha})t}=\gamma_{j}\delta(t),
\end{eqnarray}
we obtain the following relations after some algebra

\begin{eqnarray}
Tr_{B}\int_{0}^{\infty}dt'V(t)V(t')\rho_{s}(t')\otimes\rho_{B} &=& \sum_{j}\gamma_{j}[n_{j}(\sigma_{j}^{-}+\sigma_{j+1}^{-})(\sigma_{j}^{+}
+\sigma_{j+1}^{+})+(n_{j}+1)(\sigma_{j}^{+}+\sigma_{j+1}^{+})(\sigma_{j}^{-}+\sigma_{j+1}^{-})]\rho_{s}, \\
Tr_{B}\int_{0}^{\infty}dt'V(t)\rho_{s}(t')\otimes\rho_{B}V(t') &=&
\sum_{j}\gamma_{j}[(n_{j}+1)(\sigma_{j}^{-}+\sigma_{j+1}^{-})\rho_{s}(\sigma_{j}^{+}+\sigma_{j+1}^{+})
+n_{j}(\sigma_{j}^{+}+\sigma_{j+1}^{+})\rho_{s}(\sigma_{j}^{-}+\sigma_{j+1}^{-})],\\
Tr_{B}\int_{0}^{\infty}dt'V(t')\rho_{s}(t')\otimes\rho_{B}V(t) &=&
\sum_{j}\gamma_{j}[(n_{j}+1)(\sigma_{j}^{-}+\sigma_{j+1}^{-})\rho_{s}(\sigma_{j}^{+}+\sigma_{j+1}^{+})
+n_{j}(\sigma_{j}^{+}+\sigma_{j+1}^{+})\rho_{s}(\sigma_{j}^{-}+\sigma_{j+1}^{-})],\\
Tr_{B}\int_{0}^{\infty}dt'\rho_{s}(t')\otimes\rho_{B}V(t')V(t) &=&
\sum_{j}\gamma_{j}\rho_{s}[n_{j}(\sigma_{j}^{-}+\sigma_{j+1}^{-})(\sigma_{j}^{+}
+\sigma_{j+1}^{+})+(n_{j}+1)(\sigma_{j}^{+}+\sigma_{j+1}^{+})(\sigma_{j}^{-}+\sigma_{j+1}^{-})].
\end{eqnarray}
\end{widetext}
In Eq.(\ref{reduced_density_matrix}), the first term $-i Tr_{B}[V(t),\rho_{s}(0)\otimes\rho_{B}]$ is zero due to
the properties of thermal bath $\langle b_k \rangle = \langle b^{\dagger}_k \rangle = 0$.
Combing the above equations and Eq.(\ref{reduced_density_matrix}), we can get the master equation of
Eq.(\ref{master_equation}).

\end{document}